%% file: embright_paper.tex
\documentclass[twocolumn]{aastex63}

\usepackage{graphicx}
\usepackage{amsmath, amssymb}
\usepackage{placeins}
\usepackage{booktabs}
\usepackage{units}
\usepackage[caption=false]{subfig}
\usepackage{float}
\usepackage[caption=false]{subfig}

\newcommand{\bbeta}{\pmb{\beta}}
\newcommand{\hasns}{p(\texttt{HasNS})}
\newcommand{\hasremnant}{p(\texttt{HasRemnant})}

\begin{document}
\title{A Machine Learning Based Source Property Inference for Compact Binary Mergers}

\author[0000-0003-0038-5468]{Deep Chatterjee}
\affiliation{Department of Physics, University of Wisconsin--Milwaukee, Milwaukee, WI 53211, USA}

\author[0000-0001-9901-6253]{Shaon Ghosh}
\affiliation{Department of Physics, University of Wisconsin--Milwaukee, Milwaukee, WI 53211, USA}
\affiliation{Department of Physics and Astronomy, Montclair State University, 1 Normal Avenue, Montclair, NJ 07043, USA}

\author[0000-0002-4611-9387]{Patrick R. Brady}
\affiliation{Department of Physics, University of Wisconsin--Milwaukee, Milwaukee, WI 53211, USA}

\author[0000-0001-5318-1253]{Shasvath J. Kapadia}
\affiliation{Department of Physics, University of Wisconsin--Milwaukee, Milwaukee, WI 53211, USA}
\affiliation{International Centre for Theoretical Sciences, Tata Institute of Fundamental Research, Bangalore 560012, India}

\author{Andrew L. Miller}
\affiliation{Centre for Cosmology, Particle Physics and Phenomenology (CP3),
Universit\'{e} Catholique de Louvain, Chemin du Cyclotron, 2\\B-1348 Louvain-la-Neuve, Belgium}

\author{Samaya Nissanke}
\affiliation{GRAPPA, Anton Pannekoek Institute for Astronomy and Institute of High-Energy Physics, University of Amsterdam, Science Park 904,
1098 XH Amsterdam, The Netherlands}

\author[0000-0002-7537-3210]{Francesco Pannarale}
\affiliation{Dipartimento di Fisica, Universit\`{a} di Roma ``Sapienza,'' Piazzale A. Moro 5, I-00185 Roma, Italy}
\affiliation{INFN Sezione di Roma, Piazzale A. Moro 5, I-00185 Roma, Italy}

\begin{abstract}
\input{abstract}
\end{abstract}

\section{Introduction}
\label{sec:intro}
\input{introduction}

\section{Ellipsoid based classification}
\label{sec:ellipsoid_based_classification}
\input{ellipsoid_based_classification.tex}

\section{Machine learning based classification}
\label{sec:ml_based_classification}
\input{ml_based_classification.tex}
\section{Conclusion}
\label{sec:conclusion}
\input{conclusion}

\acknowledgments
\input{acknowledgements}

\appendix
\input{appendix}

\bibliography{references}

\end{document}

%% file: abstract.tex
The detection of the binary neutron star (BNS) merger, GW170817, was
the first success story of multi-messenger observations
of compact binary mergers. The inferred merger rate along with
the increased sensitivity of the ground-based gravitational-wave (GW)
network in the \replaced{third}{present} LIGO/Virgo, \added{and future LIGO/Virgo/KAGRA} 
observing runs, strongly hints at detections
of binaries which could potentially have an electromagnetic (EM) counterpart.
A rapid assessment of properties that could lead to a counterpart is
essential to aid time-sensitive follow-up operations, especially robotic
telescopes.
At minimum, the possibility of counterparts require
a neutron star (NS). Also, the tidal disruption physics is important to determine
the remnant matter post merger, the dynamics of which could result in the counterparts.
The main challenge, however, is that the
binary system parameters such as masses and spins estimated from the realtime,
GW template-based searches are often dominated by statistical
and systematic errors. Here, we present an approach that uses supervised machine-learning
to mitigate such selection effects to report possibility of counterparts
based on presence of a NS component, and presence of remnant matter post merger
in realtime.

%% file: introduction.tex
The first two observing runs of the LIGO detectors, \citep{advanced_ligo}
and the Virgo detector \citep{Acernese_2014} witnessed
remarkable level of participation from the electromagnetic
(EM) astronomy community in search for EM counterparts of gravitational
wave (GW) detections from coalescing binaries \citep{ligo_O2_emfollow_2019, ligo_catalog_paper}.
As the detectors become more sensitive, the projected detection rates of
such events will increase \citep{ligo_2018_obs_scenario}.
Technological improvement is not just confined to GW
detectors alone. Current and upcoming telescope facilities such as the
Zwicky Transient Facility \citep{ztf_kulkarni} and the Large Synoptic Survey
Telescope, \citep{lsst} consistent with the timeline of LIGO/Virgo operations,
plan to participate in the follow-up efforts (see \cite{Graham_2019}, for example).

Observers are interested to know about the presence of a neutron
star (NS) in coalescing binaries. 
This is a minimum condition for there
to be matter post merger. The dynamics of matter in the extreme
environment of the aftermath of a compact binary merger is responsible for EM
phenomena associated with GWs. Binary black hole (BBH)
mergers, therefore, are not expected to have an associated counterpart,
since they are vacuum solutions to the Einstein's field equations.
Even in the presence of a NS, other effects, like the equation of state
(EoS) of the NS(s), or the mass and spin of the companion BH plays crucial
role in the tidal disruption, and the amount of matter ejected. For a neutron
star black hole (NSBH) system, tidally disrupted material from the NS could
form an accretion disk around the central BH. High temperatures in the
disk could lead to annihilation of neutrinos to pair produce electron-positrons,
which further annihilate to power a short GRB. This could also happen via
extraction of rotational energy from the BH due to the presence of magnetic
field lines threading the BH horizon \citep{blandford_znajek}.
In the case of unbound ejecta,
$r$-process nucleosynthesis can power a \emph{kilonova}.
\citep{lattimer_1974, Li_1998, korobkin_2012, Tanaka_2013, Barnes_2013,
kasen_2015} For a binary neutron star (BNS) system, even if
the tidal interaction is not strong enough, the two bodies will eventually
come into physical contact, resulting in shocks that
expel neutron rich material. This will result in a kilonova as seen
in the case of GW170817 \citep{mma_2017, arcavi_2017, Coulter_2017,
kasliwal_2017, Lipunov_2017, Soares_Santos_2017, Tanvir_2017}.
The interaction of the ejecta with the surrounding
medium can result in synchrotron emission, observable in X-rays and
radio in weeks to months. There can be relativistic outflows, which could result in
a GRB, as seen for GW170817; although, there could be cases of prompt
collapse where GRB generation could be suppressed \citep{Ruiz_2017}.
Nevertheless, the generation of some EM messenger is highly probable.
Therefore, data products that predict the existence of matter is
useful in the EM counterpart follow-up operations.

An accurate computation of the remnant matter requires general-relativistic
numerical simulations of compact mergers. These are expensive, and
only a few ($\lesssim 100$) such simulations have been performed to date.
Also, such a simulation is not possible in the time scale of discovery,
and generic target of opportunity follow-up of GW candidates. Empirical
fits to the numerical relativity results, however, have been performed,
and are a use case for such realtime inferences. For example, 
\cite{PhysRevD.86.124007} and \cite{PhysRevD.98.081501} devised an empirical
fit to predict the combined mass from
the accretion disk, the tidal tail, and the ejecta remaining
outside the final BH in case of a NSBH merger. However,
it should be mentioned that such fits often require more input than what
is available from the realtime GW data. For example, the fits mentioned
above require the compactness of the NS, which is not a parameter inferred
by the GW searches. The NS EoS, which is not constrained strongly, is to be
assumed in order to infer the compactness.

The second LIGO/Virgo observing run, O2, saw the first effort to provide realtime
data products to aid EM follow-up operations from ground and
space based facilities \citep{ligo_O2_emfollow_2019}. These included sky
localization maps, \citep{PhysRevD.93.024013, 2041-8205-829-1-L15} and
source classification of the binary which included
\begin{enumerate}
    \item the probability that there was at least one neutron star in
          the binary, $\hasns$, and
    \item the probability that there was non-zero remnant matter,
          $\hasremnant$, considering the mass and spin of the components,
          based on the \cite{PhysRevD.86.124007} fit.
\end{enumerate}
For a BNS merger, we expect some matter to be expelled (see Table 1 of
\cite{shibata_review_2019} for different scenarios). Therefore, we expect
the result, $\hasns = 1$; $\hasremnant = 1$. On the other
extreme, BBH coalescences will not lead to remnant matter,
since they are vacuum solutions, i.e., $\hasns = 0$; $\hasremnant = 0$.
Hence, $\hasremnant$ is more relevant for NSBH systems.
Here, the mass and spin of the BH determines the tidal
disruption of the NS. Lower mass, and high spin implies a smaller
innermost stable circular orbit which allows the NS to inspiral
closer to BH. The tidal force exerted by the BH, which also increases
with spin, then tears the NS apart. This leaves remnant matter post
merger. However, if the NS is compact, or tidal forces are not sufficient
enough, the NS is swallowed whole into the BH, leaving no remnant.
The type and morphology of EM counterparts generated depends
on the amount of matter ejected and its properties. \cite{Pannarale_2014}
considered the conditions for short GRB production in the context
of LIGO/Virgo observations of NSBHs. More recent work has tried to understand
the morphology of kilonovae from NSBH mergers considering the 
density structure of the ejected matter, opacity properties, the
viewing angle, and other factors (see \cite{Barbieri_2019, hotokezaka2019radioactive},
for example). However, accurate modeling is still at its infancy.
Thus, the presence of remnant matter is a conservative proxy for the
presence of counterparts, still more constraining than the presence
of a NS component alone, albeit the model dependence i.e., the assumption
of NS EoS, and the usage of a particular fit. The rationale behind
computing two quantities is to give flexibility to observing partners
in follow-up operations.

The main challenge in this inference, however, is to handle detection
uncertainties in the parameter recovery of the realtime GW template-based searches.
This was done in O2 via an effective Fisher formalism using an \emph{ambiguity}
region around the parameters of the triggered template. The algorithm used for
O2 is described in Sec. 3.3.2 of \cite{ligo_O2_emfollow_2019}, and briefly summarized in
Sec.~\ref{sec:ellipsoid_based_classification} below. While it accounted for
statistical uncertainties, the systematic errors in the low-latency GW template
based analysis were not considered. Here we consider the problem differently.
We treat the problem as binary classification, and present a new technique
that is based on supervised learning. This not only improves the speed and
accuracy, but also removes runtime dependencies that were required during O2
operations. Also, this technique provides flexibility to incorporate astrophysical
rates of binary populations in the universe.

In the third LIGO/Virgo observing run, O3, these data products (and a few more)
continue to be part of the public alerts. \footnote{
\url{https://emfollow.docs.ligo.org/userguide/}}
In this work, we make a slight modification to the nomenclature.
The $\hasremnant$ quantity had been referred to as \emph{EM bright} classification
probability in \cite{ligo_O2_emfollow_2019}. Here, we refer to the both these
quantities collectively as \textit{source properties}, following the O3 LIGO/Virgo
public alert userguide. These values indicate the chances of the matter remaining
post merger, the dynamics of which can launch EM counterparts. 
For example, the combination {$\hasns=1$; $\hasremnant=0$}, indicates a
conservative measure of presence of matter -- just the presence of NS. However,
the combination {$\hasns=1$; $\hasremnant=1$}, is a stronger indication of the
presence of a counterpart, albeit some model dependence.

The organization of the paper is as follows. In Sec.~\ref{sec:ellipsoid_based_classification}
we provide a brief review of the ellipsoid-based inference used in O2.
In Sec.~\ref{sec:ml_based_classification}, we present the inference using a supervised
learning method called \texttt{KNeighborClassifier}
\citep{scikit-learn}, which was trained on injection campaigns from the GstLAL
search pipeline \citep{gstlal_2017} used by LIGO/Virgo in routine
search sensitivity analyses during O2. We test the performance
of the machine learned inference. In Sec.~\ref{sec:conclusion}, we
conclude and propose to use this method to report source properties,
$\hasns$ and $\hasremnant$ in future operations.

%% file: ellipsoid_based_classification.tex
\subsection{Low-latency Searches}
\begin{figure*}[htp]
	\begin{center}
	    \includegraphics[width=0.44\textwidth]
	        {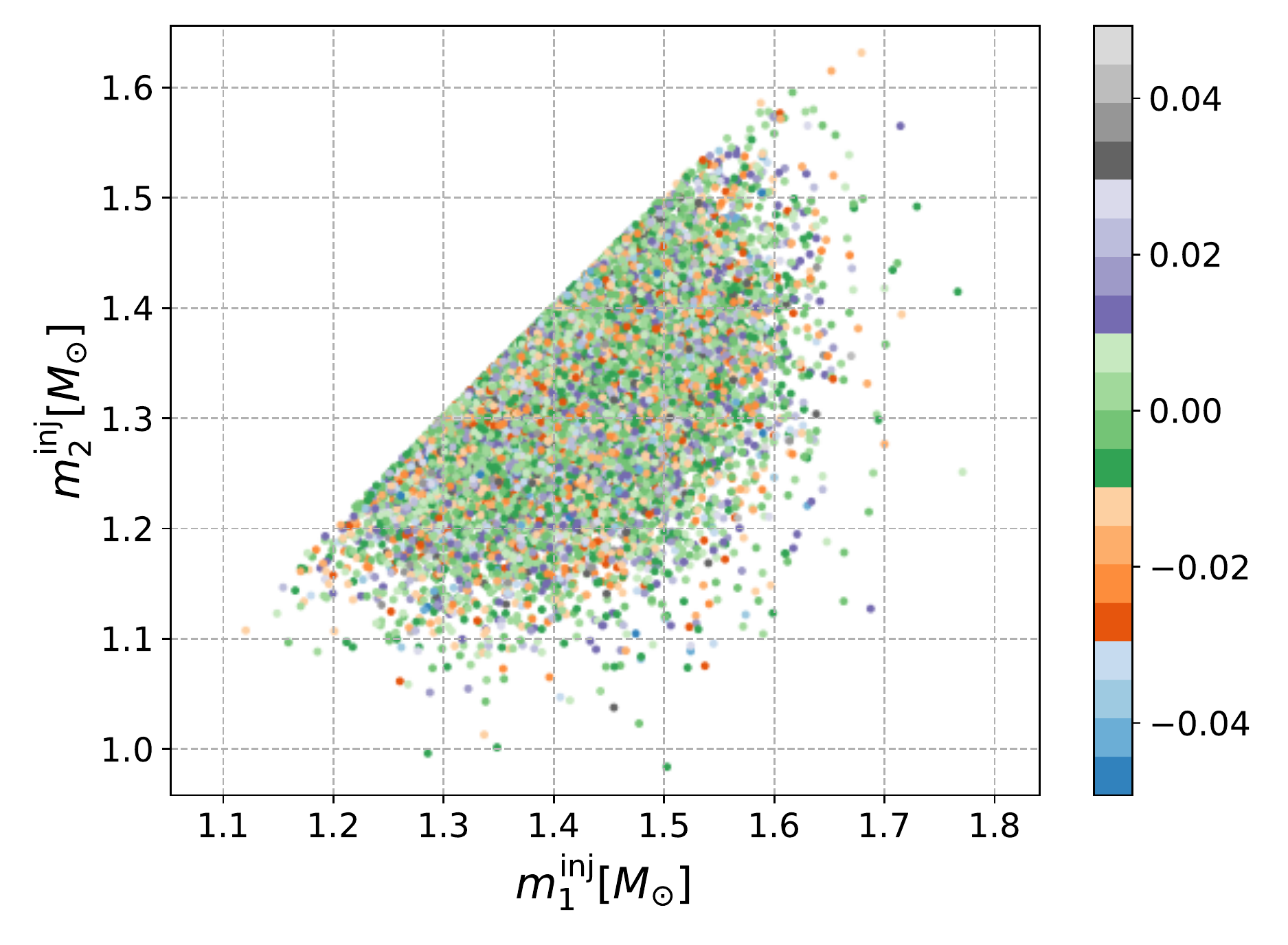}
	    \includegraphics[width=0.44\textwidth]
	        {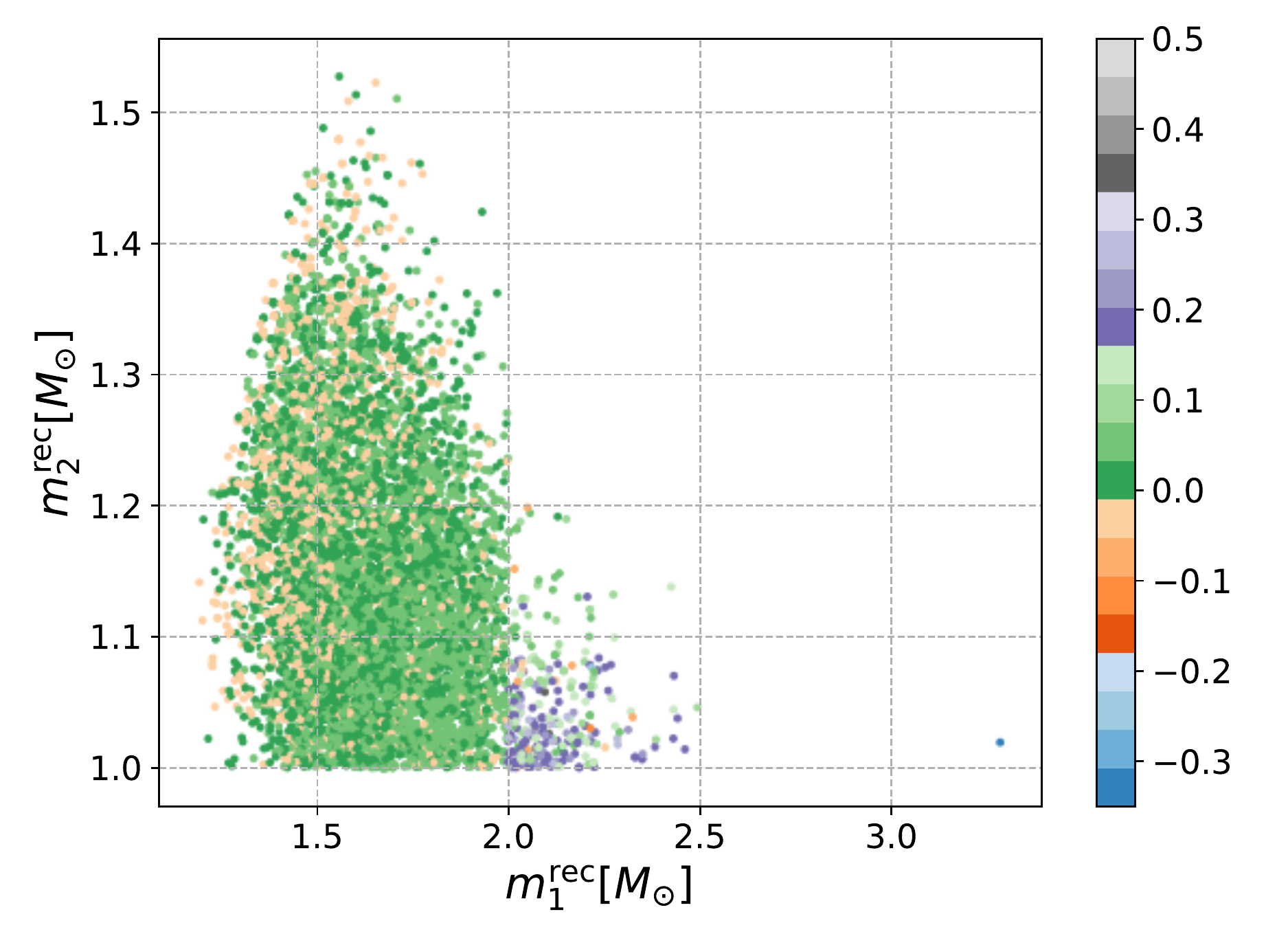}
	    \includegraphics[width=0.44\textwidth]
	        {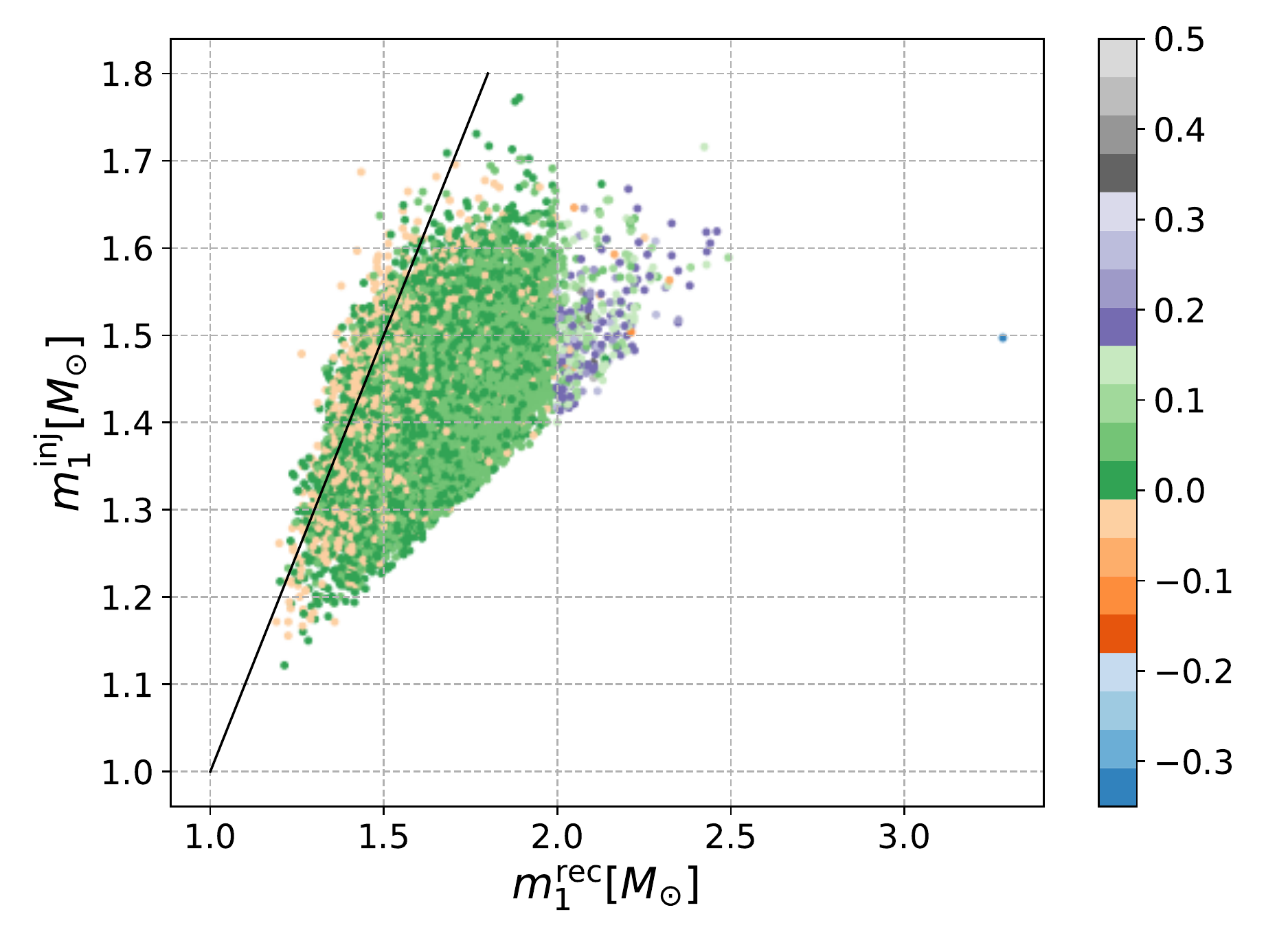}
	    \includegraphics[width=0.44\textwidth]
	        {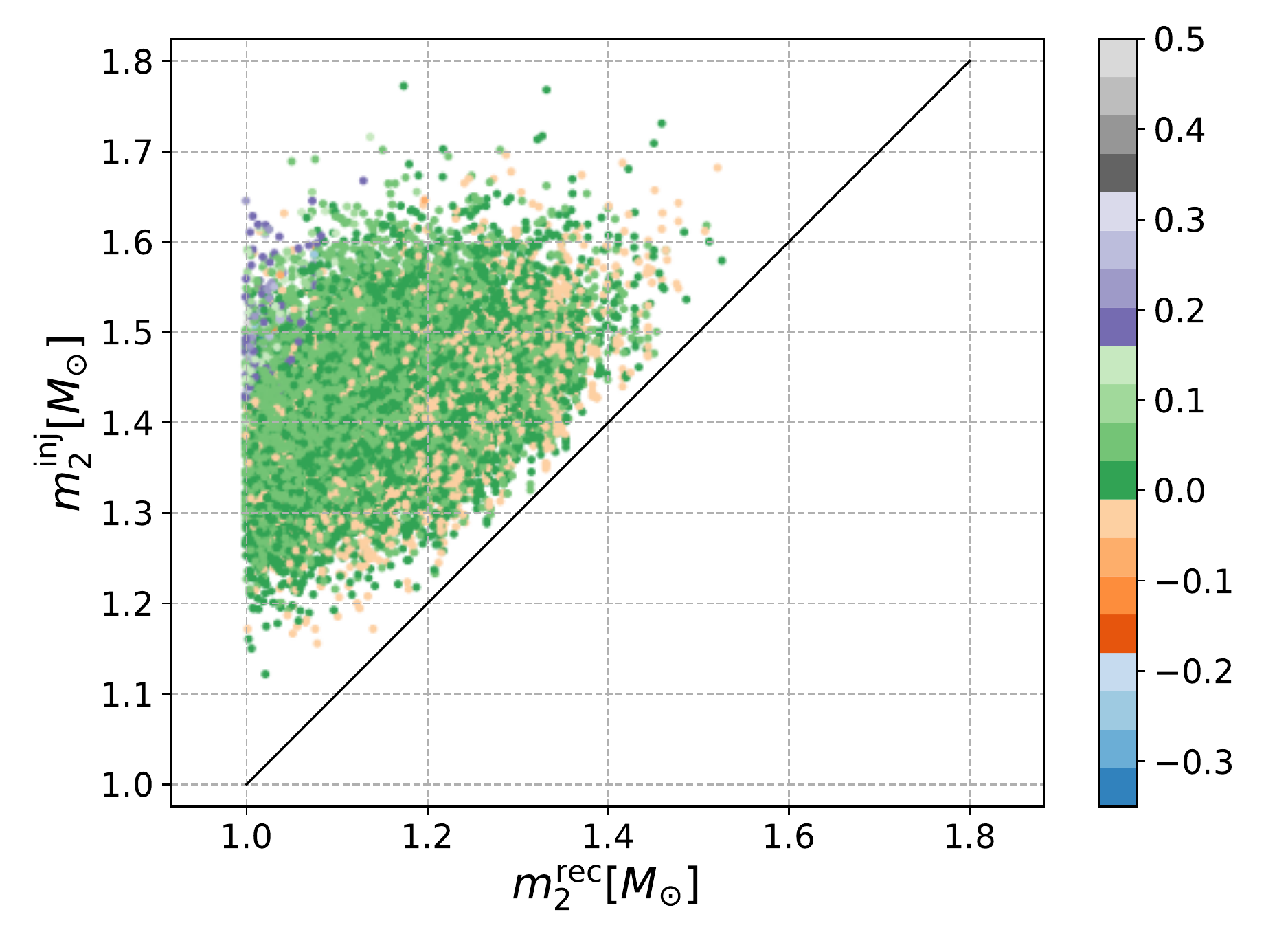}
	\end{center}
	\caption{
    In this figure we compare the mass and spin recovery of one of
    the search pipelines, GstLAL \citep{gstlal_2017}, that meet the
    false alarm rate threshold of Eq.(\ref{eq:ml_far_threshold}).	
	\textbf{Upper panel}: This panel shows the $(m_1, m_2)$ pairs of
	a gaussian distributed BNS population $\sim \mathcal{N}[1.33 M_{\odot}, 0.09 M_{\odot}]$
	(see Table~\ref{tab:ml_population_distribution}). The left plot shows the
	masses injected following a normal distribution, as mentioned in
	Table~\ref{tab:ml_population_distribution}, colored by the injected
	primary aligned spin component, $\chi_1^z$. The right plot
	shows the recovered masses colored by the recovered $\chi_1^z$.
	It can be seen that the distribution in the recovered space is
	significantly different from the one in the injected space. One may
	also see that the recovered spin values may be higher than
	the injected ones, especially in the case of higher
	mass ratio recoveries.
	\textbf{Lower panel}: This panel shows the injected values of
	the primary and secondary masses against their recovered values
	for low-mass injections. This is an example where one can see the
	systematic effect of the primary mass being recovered at higher
	values than the injected values. The secondary follows the
	opposite trend: the recovered value is lesser than the injected values.
	The effect also exists at higher mass ranges. Both plots are colored
	by the recovered $\chi_1^z$ values.
	Note the recovered $m_1^{\text{rec}} > 2 M_{\odot}$ (both panels)
	have higher values of recovered $\chi_1^z$. This is because the
	GstLAL search uses templates with low spins for masses
	$\leq 2 M_{\odot}$ and high spins above that (see Fig. 1 $\&$ 2 in
	\cite{2018arXiv181205121M} for example). Even values slightly higher
	than $2 M_{\odot}$ may result in high spin values compared to
	the injections.
	}
	\label{fig:ellipsoid_DetPipelineRec}
\end{figure*}

LIGO/Virgo searches for transient GW signals fall into two broad categories: modeled compact
binary coalescence (CBC) searches \citep{mbta_2016, gstlal_2017, chu_2017, pycbc_2018, ligo_catalog_paper}
and un-modeled burst searches \citep{PhysRevD.95.104046, PhysRevD.93.042004}.
In this work, we are concerned with the former. The modeled searches
use a discrete
template bank of CBC waveforms to carry out 
matched filtering on the data. This is further broken down
into realtime online analysis, and calibration corrected offline analysis. The
online low-latency searches report CBC events in sub-minute latencies. They use
waveform templates that are characterized by masses, $(m_1, m_2)$, and the
dimensionless aligned/anti-aligned spins of the binary elements along the orbital
angular momentum of the binary, $(\chi_1^z, \chi_2^z)$. They report a best matching
template based on an appropriate detection statistic. We call the parameters of
this template, $\{m_1, m_2, \chi_1^z, \chi_2^z\}$, the \emph{point estimate}. This
data can be used for low-latency source property inference.

\subsection{Capturing Detection Uncertainties}
Since the source property inference is to be done based on the
point estimates, the obvious pitfall in the inference is: How accurate are the
point-estimates compared to the true parameters of the source? The primary goal
of detection pipelines is to maximize detection efficiency at fixed false alarm
probability. While some parameters like the chirp mass,
\begin{equation}
	\mathcal{M}_c = (m_1 m_2)^{3/5}/(m_1 + m_2)^{1/5},
	\label{eq:ml_mchirp_def}
\end{equation}
on which the signal strongly depends, are measured accurately,
\footnote{More precisely, this is true for low-mass systems
where the waveform is dominated by the inspiral phase. For heavier
BBH systems, the total mass, $m_1 + m_2$, is recovered accurately.}
others like the individual mass or spin components are often
inconsistent compared to the true parameters. Accurate parameter
recovery is left to Bayesian parameter estimation analysis
\citep{PhysRevD.91.042003, bilby, pycbc_inference}.

Consider the case for the GstLAL search \citep{gstlal_2017, 2018arXiv181205121M,
gstlal_2019} in Fig.~\ref{fig:ellipsoid_DetPipelineRec}. Here,
we compare fake GW signals whose parameters we know a priori, to
the recovered template i.e., point estimate, obtained from injecting
the fake signals in detector noise and running the pipeline.
Note that the recovered masses can sometimes be significantly different
from the injected values, leading to an erroneous classification of the
systems based on point-estimates alone. To alleviate this problem
attempts were made to capture the uncertainty in the recovery of
the parameters using an effective Fisher formalism \citep{PhysRevD.87.024004}.
This method allows us to construct an ellipsoidal region of the parameter
space around the point estimate that captures the uncertainty in
the parameters under the Fisher approximation. This was
used to create confidence regions in the parameter estimation code,
\texttt{RapidPE} \citep{pankow_2015} from which it was implemented in
\texttt{EM-Bright} pipeline to construct $90\%$ confidence regions
in three dimensions -- chirp mass, symmetric mass ratio and effective spin.
This ellipsoidal region was populated uniformly with one thousand points (besides the
original triggered point). The fraction of these ellipsoid samples which had
$m_2 < m^{\text{NS}}_{\text{max}}$\footnote{$m^{\text{NS}}_{\text{max}} = 2.83 M_{\odot}$ was used during
O2 operations. This is the maximum allowed mass of a NS assuming the 2H
EoS.}
constituted the $\hasns$ value, while the fraction that had non-vanishing
disk mass, $M_{\text{disk}} > 0$ from the \cite{PhysRevD.86.124007} fit,
constituted $\hasremnant$ value.

%% file: ml_based_classification.tex
\begin{deluxetable*}{llll}
    \tablecaption{The table lists the different population
    features used in the injection campaign. This includes
    signals in the three categories of CBC signals - binary
    black hole (BBH), neutron star black hole (NSBH) and
    binary neutron star (BNS) categories. The BBH category
    has both aligned and isotropic spin distributions.
    The BNS category has high spinning and low spinning
    systems to account for isolated high spinning neutron
    stars and galactic binaries. The NSBH category, includes
    $\delta$ function distributions along with uniform in log
    mass distribution. The $U, \mathcal{N}, \delta$ imply
    uniform, normal and delta function distributions respectively.
    These injections densely sample possible populations of binaries. The
    number of found injections that passed the FAR threshold in
    Eq.~(\ref{eq:ml_far_threshold}) used in training are listed in the
    right-most column. The campaign uses the SpinTaylorT4 approximant
    for BNS injections, and effective one body calibrated to numerical
    relativity SEOBNR approximant for NSBH and BBH injections.
    }
    \tablehead{
    \colhead{Type} & \colhead{Mass distribution} &
    \colhead{Spin distribution} & \colhead{Num. Injections}
    }
    \startdata
    BBH & $U[\log m_1, \log m_2]$ & $\vert\chi^{\mathrm{max}}\vert = 0.99$ (Isotropic) & $4.0\times 10^4$ \\
    BBH & $U[\log m_1, \log m_2]$ & $\vert\chi^{\mathrm{max}}\vert = 0.99$ (Aligned) & $1.9 \times 10^4$ \\
    BNS & $\mathcal{N}[1.33 M_{\odot}, 0.09 M_{\odot}]$
            & $\vert\chi^{\mathrm{max}}\vert = 0.05$ (Isotropic) & $1.6\times 10^4$ \\
    BNS & $U[m_1, m_2]$ &
            $\vert\chi^{\mathrm{max}}\vert = 0.40$ (Isotropic) & $1.6\times 10^4$ \\
    NSBH & $U[\log m_1, \log m_2]$ &
    		$\vert\chi^{\mathrm{max}}_{\mathrm{NS}}\vert = 0.40$;
            $\vert\chi^{\mathrm{max}}_{\mathrm{BH}}\vert = 0.99$ (Aligned) & $1.9\times 10^4$ \\
    NSBH & $\delta(m_1 - 5 M_{\odot}, m_2 - 1.4 M_{\odot})$ &
            $\vert\chi^{\mathrm{max}}_{\mathrm{NS}}\vert = 0.05$;
            $\vert\chi^{\mathrm{max}}_{\mathrm{BH}}\vert = 0.99$ (Aligned/Isotropic) & $1.6\times 10^4$/$1.5\times 10^4$ \\
	NSBH & $\delta(m_1 - 10 M_{\odot}, m_2 - 1.4 M_{\odot})$ &
            $\vert\chi^{\mathrm{max}}_{\mathrm{NS}}\vert = 0.05$;
            $\vert\chi^{\mathrm{max}}_{\mathrm{BH}}\vert = 0.99$ (Aligned/Isotropic) & $1.7\times 10^4$/$1.3\times 10^4$ \\
	NSBH & $\delta(m_1 - 30 M_{\odot}, m_2 - 1.4 M_{\odot})$ &
            $\vert\chi^{\mathrm{max}}_{\mathrm{NS}}\vert = 0.05$;
            $\vert\chi^{\mathrm{max}}_{\mathrm{BH}}\vert = 0.99$ (Aligned/Isotropic) & $1.8\times 10^4$/$1.3\times 10^4$ \\
    \enddata
\end{deluxetable*}\label{tab:ml_population_distribution}
\begin{figure}[h]
    \begin{center}
    \includegraphics[width=0.47\columnwidth]{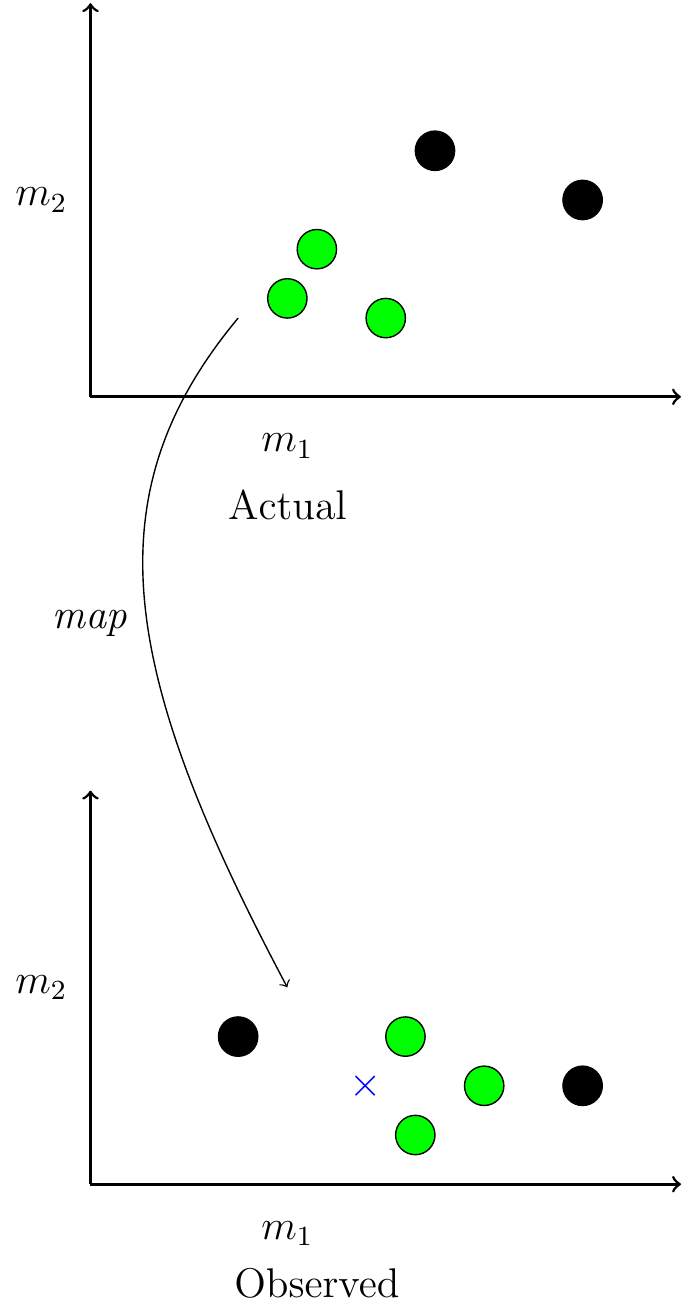}
    \end{center}
    \caption{This figure is an qualitative illustration of the binary
    classification treatment of the problem. The top panel represent
    the true parameter space of binaries i.e, the injected parameters
    in this case, where the two colors represents satisfying either of
    the conditions in Eq.~(\ref{eq:ml_hasns_label},
    \ref{eq:ml_hasremnant_label}). The lower panel is the parameter space
    of the recovery i.e., what the search reports. For the training
    process, the parameters in the recovered space are the features, while
    the label is inferred from the actual parameters. A fiducial detection
    during the production running is represented by the $\times$ mark in this
    plane. The probability of this fiducial detection being either of the
    two binary classes is determined from the nearest neighbors in the recovered
    parameter space.}
    \label{fig:knn_illustration}
\end{figure}
The method of uncertainty ellipsoids handles the statistical uncertainties of
the parameters from the low-latency search pipelines. However, the underlying
Fisher approximation is only suitable in the case of high signal to noise ratio,
when the parameter uncertainties are expected to be Gaussian distributed (see
Sec. II of \cite{PhysRevD.49.2658} for example). Also, it is not robust
in capturing any bias that a search might have. Such trends are seen, for example,
in Fig.~\ref{fig:ellipsoid_DetPipelineRec} where the $m_1$ parameter is
recovered to be larger than the injected value, while the $m_2$ parameter is
recovered to be smaller.\footnote{In GW parameter estimation, $m_1$ refers to the
primary (larger) mass component while $m_2$ refers to the secondary (smaller) mass
component. Likewise, $\chi_1^z$ ($\chi_2^z$) refers to the aligned spin component
of the primary (secondary).}
Such uncertainties are more often the dominant source of error in this inference.
While they decrease as the significance increases, they may be pronounced otherwise.
Capturing and correcting such selection effects can be done by supervised machine
learning algorithms. By injecting fake signals into real noise, performing the
search, and comparing the recovered parameters with the original parameters of
injections, one gets the \emph{map} between the injected and recovered parameters.
This is qualitatively illustrated in Fig.~\ref{fig:knn_illustration}.
Given a broad training set, the supervised algorithm learns this map. The
training features are recovered parameters obtained after running the search,
however, the labels of having a NS or remnant are determined from the injected
values. It should be highlighted that we are not using machine learning to predict
the recovered parameters from the injected values, or vice versa. Rather we use
it for binary classification, correcting
for selection biases that could have, otherwise, given an erroneous answer from the point
estimate. We return the probability that the binary had a component
less that $3 M_{\odot}$, which we assume to be a conservative upper limit of the
NS mass, and the probability that it had remnant matter based on the
\cite{PhysRevD.98.081501} (hereafter F18) expression.

\subsection{Injection Campaign}
In this study, we use a broad injection set that well samples the space of
compact binaries. The distribution of the masses and spins is
tabulated in Table~\ref{tab:ml_population_distribution}. The injections
are simulated waveforms placed in real detector noise at specific times.
The BNS injections use the SpinTaylorT4 approximant \citep{PhysRevD.80.084043},
while NSBH and BBH injections use the SEOBNR approximant \citep{PhysRevD.95.044028}.
We consider the injections made in two detector operations from O2 (see
Table~\ref{tab:appendix_injection_sets} for times).
The population contains uniform/log-uniform distribution of the masses, and both
aligned and isotropic distributions of spins. It was used for the spacetime
volume sensitivity analysis for the GstLAL search in \cite{ligo_catalog_paper}.
\replaced{Such injection campaigns are routinely performed in determining the
search sensitivity and can be used (as a by-product) for training in studies
such as ours.}{In particular, injection campaigns were conducted for all astrophysical
categories (BNS, NSBH, BBH) to analyze search sensitivity. We use the results,
as a by product, to train our algorithm.\footnote{The other search in
\cite{ligo_catalog_paper}(see Sec.~VII therein), PyCBC, conducted broad campaigns
for BBH population only. The method presented here, however, can be extended
to any general CBC search given suitable training data.}}

For an injection campaign, as this one, fake GW signals are put in
real detector noise, followed by which the search is run, just as in the case
of analyzing the production data. The injections maybe recovered based on the
noise properties, and the GW intrinsic (masses and spins) and extrinsic
(distance, sky location etc.) parameters. Since we are using real data,
the dynamic variation of the power spectral density is taken into account (see
Table~\ref{tab:appendix_injection_sets} for the stretch of data used, and
the splitting of the data into chunks). Not all injections are found by the
searches, partly because of the signal strength, or from having them at a
sky location where the detectors are not sensitive. The search reports triggers
coincident signal across multiple detectors, simultaneously getting a high
detection statistic. The triggers are assigned a false alarm rate (FAR)
based on the frequency of background triggers that are assigned an equal or
more significant value of the detection statistic. If the time of an injection
coincides with the time of recovery of a trigger, the injection is considered
found. For this study, we further subsample to the set where the FAR
of the recovered triggers corresponding to found injections is less
than one per month,
\begin{eqnarray}
	\text{FAR} &\leq& 1 / 1 \text{ month} \nonumber \\
	           &=& 3.85 \times 10^{-7} \text{Hz}.
    \label{eq:ml_far_threshold} 
\end{eqnarray}
This leaves us with $\sim 2.0 \times 10^5$ injections to train our
supervised algorithm. The breakdown into different populations
is shown in Table~\ref{tab:ml_population_distribution}.
This FAR threshold is reasonable
since the LIGO/Virgo public alerts in the third observing run consider
a false alarm rate threshold of one per two months further modified by
a trials factor which consider the number of independent searches
(see \url{https://emfollow.docs.ligo.org/userguide/}).

\subsection{Training Features and Performance}\label{subsec:ml_training}
\begin{figure}
    \includegraphics[width=1.0\columnwidth]
        {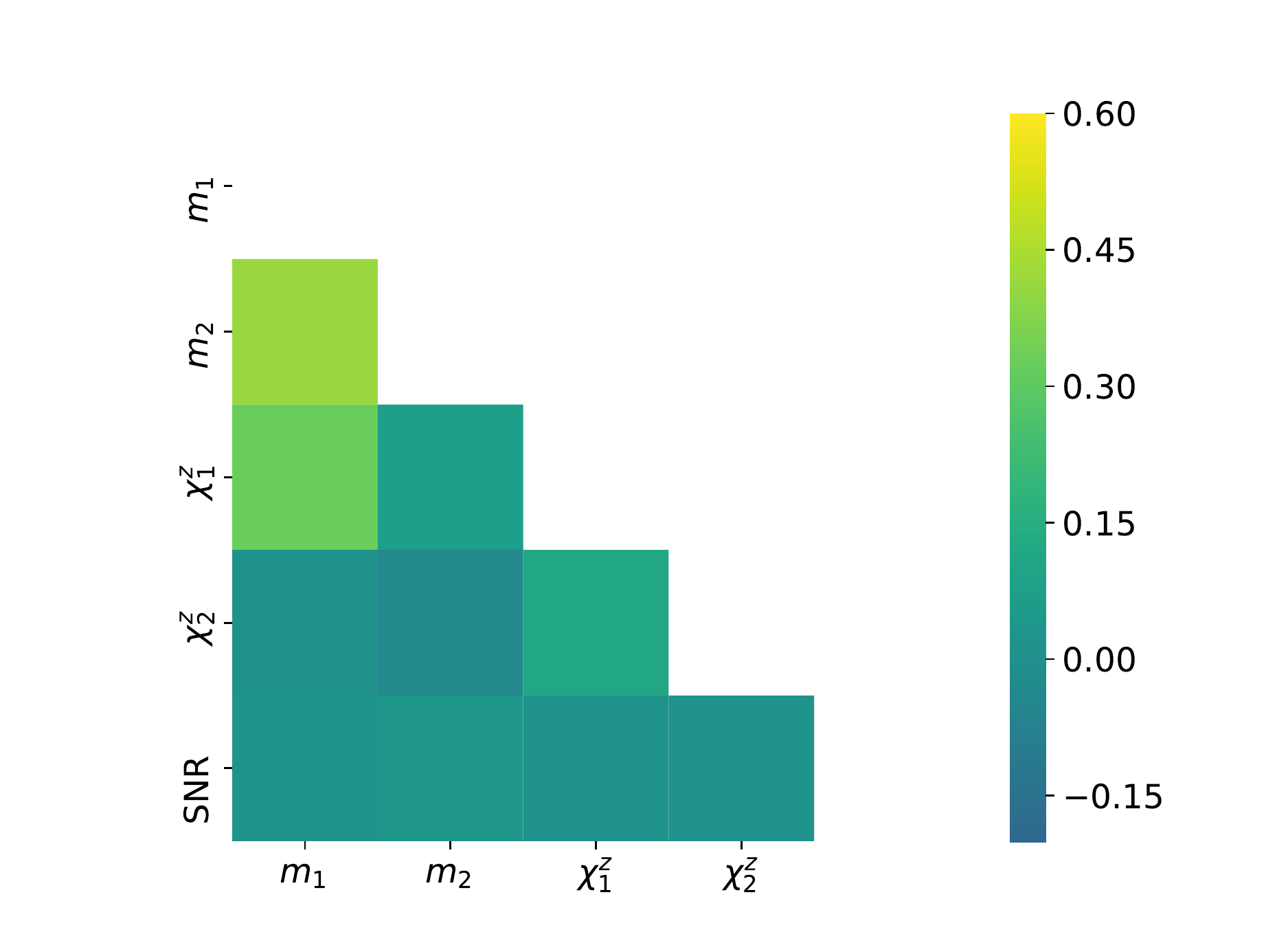}
    \caption{This is the correlation matrix of the recovered parameters
    that form our training set. The masses are expected to be correlated
    since there is a preference towards detecting heavier masses. The
    primary spin shows a strong correlation with the primary mass, however,
    the secondary spin recovery is not as correlated with the secondary mass.
    The signal-to-noise is mildly correlated with the remaining parameters.
    as expected since it is a detector frame parameter independent of the
    source properties.
    }
    \label{fig:ml_parameter_correlation}
\end{figure}
\begin{figure*}[htp]
    \begin{minipage}{0.49\textwidth}
    \begin{center}
    \includegraphics[width=1.0\textwidth]
        {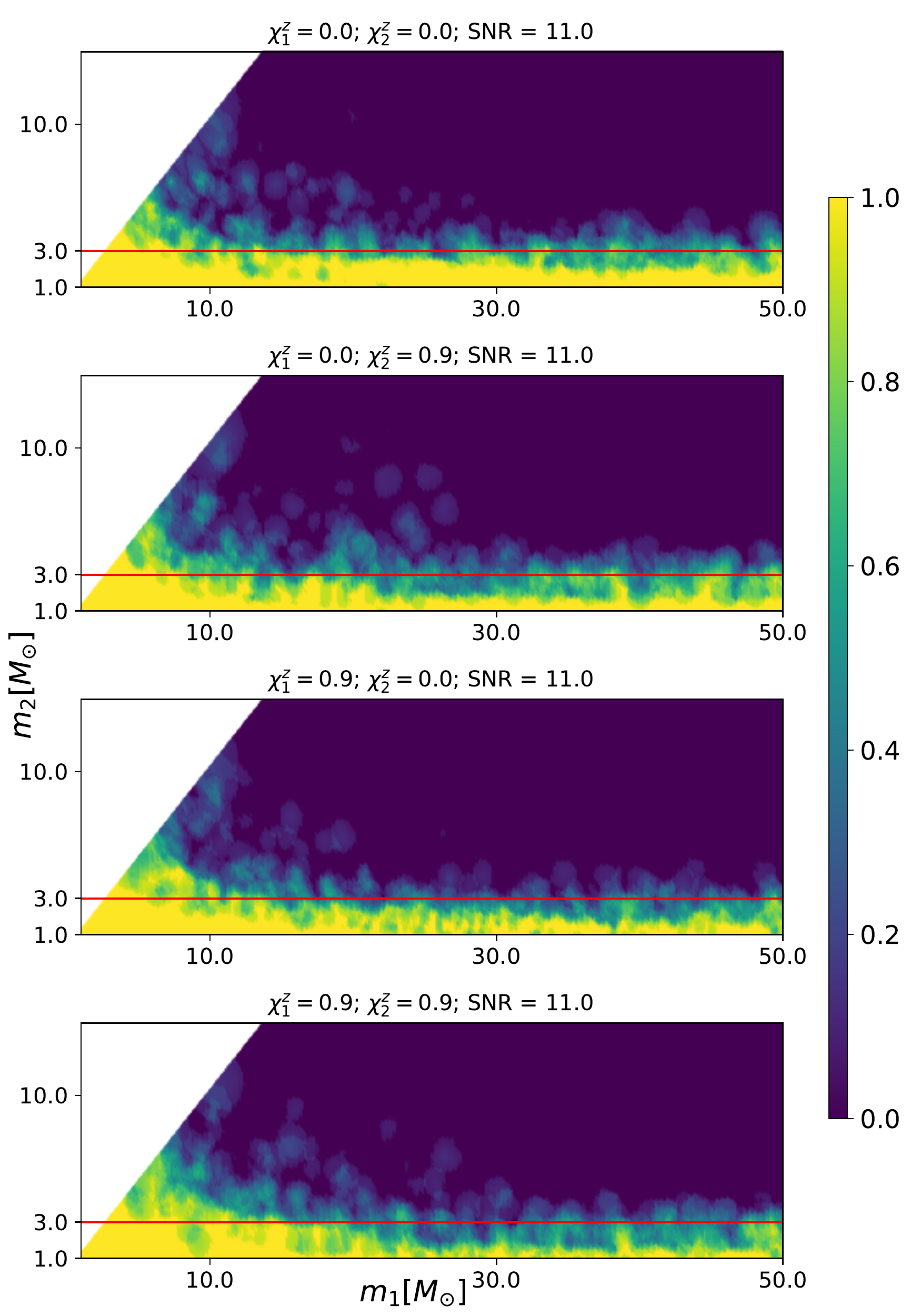} \\
    $\hasns$
    \end{center}
    \end{minipage}
    \begin{minipage}{0.49\textwidth}
    \begin{center}
        \includegraphics[width=1.0\textwidth]
        {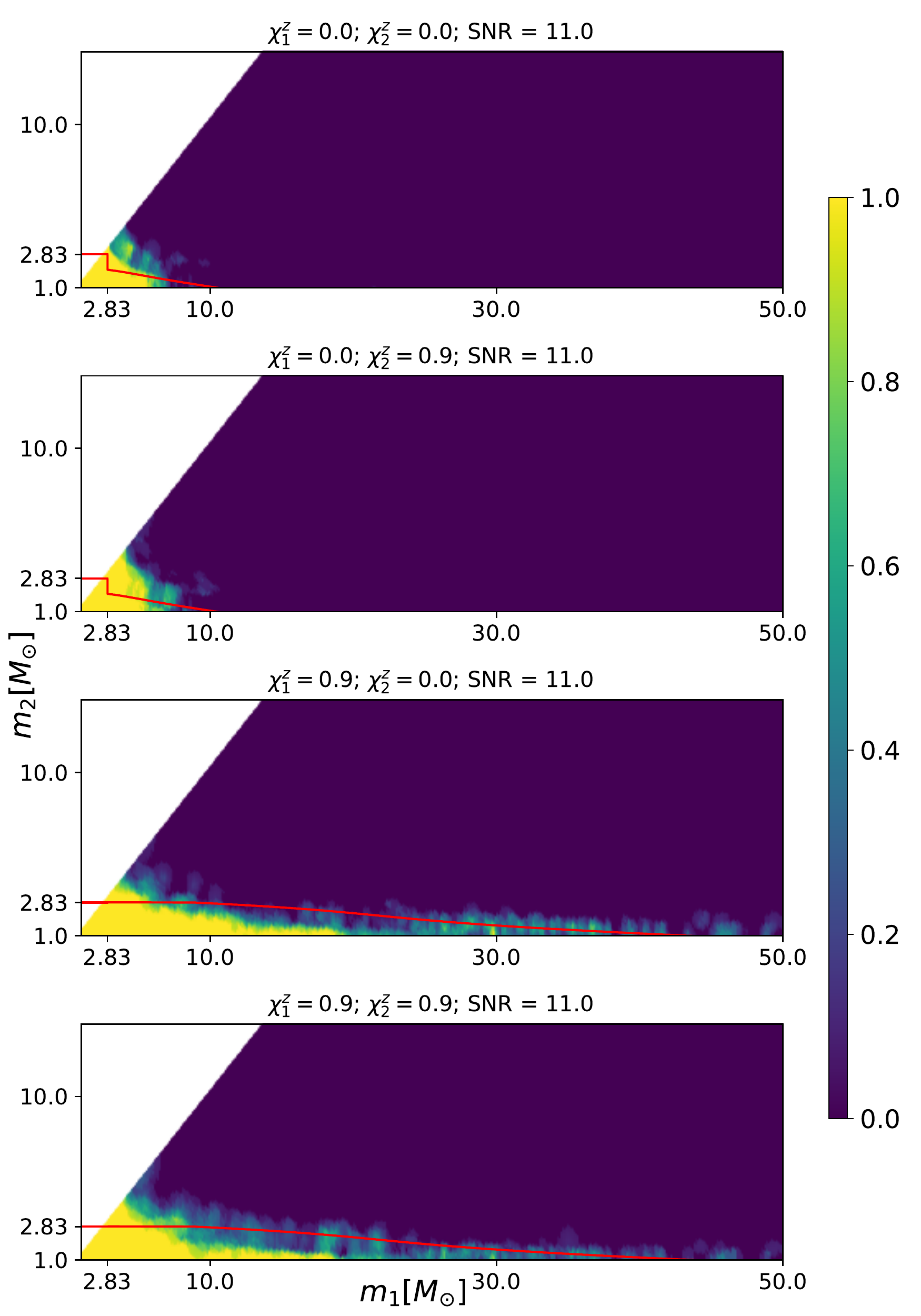} \\
    $\hasremnant$
    \end{center}
    \end{minipage}
    \caption{
        This figure shows the predictions of the \emph{trained} binary classifier
        upon performing a parameter sweep on the $(m_1, m_2)$ values.
        Note that each point on the plots is analogous to a \emph{point-estimate}.
        We feed the trained classifier with arbitrary recovered parameter
        values and evaluate the predictions. 
        \textbf{Left panel}: $\hasns$ predictions on the parameter space.
        We sweep over the masses, keeping the spin and SNR values fixed in each
        individual plot, incrementing the former as we move down. The horizontal
        line corresponds to $m_2 = 3 M_{\odot}$ around which we expect a
        \emph{fuzzy} region due to the detection uncertainties. Also,
        it is to be noted that the performance does not get affected by much upon
        increasing spin values since our original classification did not depend on it.
        Small changes are, however, expected due to correlation between the parameters
        during recovery (see Fig.~\ref{fig:ml_parameter_correlation}). \textbf{Right panel}:
        $\hasremnant$ predictions on the parameter space. The region denoting non-zero remnant
        matter shows a more constrained
        classification about presence of matter compared to
        just having a NS in the binary. Also, note that unlike $\hasns$, $\hasremnant$ is
        strongly affected by the primary spin, as expected. The red curve in this panel represents the contour
        $M_{\text{rem}}(m_1^{\text{rec}}, m_2^{\text{rec}}, \chi_1^{z {\text{ rec}}}) = 0 M_{\odot}$,
        calculated from recovered parameters using Eq.(4) of \cite{PhysRevD.98.081501}.
        Note that the $M_{\text{rem}}$ expression applies to NSBH systems and require a
        NS EoS which sets a maximum mass for the NS. In this study, we use the 2H
        EoS \citep{kyutoku_2010} which has a maximum mass of $2.83 M_{\odot}$. 
        Mass components above this maximum mass are considered BHs which do not leave
        remnant matter upon coalescence. This explains the kink in the red curve
        in the top two panels.
    }
    \label{fig:ml_parameter_sweep}
\end{figure*}
\begin{deluxetable*}{l|ccc|ccc}
    \tablecolumns{5}
    \tablecaption{The table lists the percentage misclassification when
    using a threshold of $p(\text{HasNS}/\text{HasRemnant}) = 0.5$ to
    infer a binary to have a
    counterpart, as a function of the fraction of the dataset used for
    training and testing purposes. This could be thought as the scenario when
    an external partner has decided to follow-up CBCs that report $p(\text{HasNS}) > 0.5$
    (or $p\text{HasRemnant} > 0.5$). The table lists the fraction when such an
    observation would be a false positive. Out of the fraction of the total dataset used
    (left most column), we train using $90\%$ and test on the remaining $10\%$,
    cycling the training/testing set to have predictions on all points in the set. The
    \emph{uniform} and \emph{inverse distance} weighting of the nearest neighbors
    are used in all cases. We see that the answer starts to converge when using
    $\gtrsim 50\%$ of the total dataset. In light of verifying correlations
    (shown in Fig.~\ref{fig:ml_parameter_correlation}) between parameters not affecting
    the prediction and the impurity, we trained using the \emph{Mahalanobis}
    metric \citep{mahalanobis} in the parameter space mentioned in
    Eq.(\ref{eq:ml_features}) \tablenotemark{a}. The misclassification does not
    change significantly based on the weighting scheme or the metric used.
    \tablenotemark{b}
    }
    \tablehead{
    Fraction &
    \multicolumn{3}{c|}{Misclassification $\%$ $p(\text{HasNS})$} &
    \multicolumn{3}{c}{Misclassification $\%$ $p(\text{HasRemnant})$} \\
    & Uniform & Inverse distance & Mahalanobis metric
    & Uniform & Inverse distance & Mahalanobis metric
    }
    \startdata
    $0.1$ & $3.21$ & $3.38$ & $4.24$ & $4.04$ & $4.34$ & $3.77$ \\
    $0.2$ & $3.03$ & $2.98$ & $3.80$ & $3.65$ & $3.62$ & $3.28$ \\
    $0.5$ & $2.91$ & $2.96$ & $-$ & $3.00$ & $2.92$ & $-$ \\
    $0.9$ & $2.83$ & $2.80$ & $-$ & $2.65$ & $2.64$ & $-$ \\
    $1.0$ & $2.83$ & $2.82$ & $-$ & $2.59$ & $2.59$ & $-$ \\
    \enddata
    \tablenotetext{\rm{a}}{\scriptsize{
    See \url{https://scikit-learn.org/stable/modules/generated/sklearn.neighbors.DistanceMetric.html}}
    for the implementation in the \texttt{scikit-learn} framework.
    }
    \tablenotetext{\rm{b}}{\scriptsize{
    Cross-validation when using the Mahalanobis metric is expensive and was performed for small
    fractions of the total training data.}
    }
\end{deluxetable*}\label{tab:appendix_distance_weighting}
For the $\texttt{HasNS}$ quantity, to label an injection as having a
NS, we use,
\begin{equation}
    m_2^{\text{inj}} \leq 3 M_{\odot}.
    \label{eq:ml_hasns_label}
\end{equation}
The value $\approx 3 M_{\odot}$ has been regarded as a traditional and conservative
upper limit for the NS maximum mass. The limit comes from the causality condition
of the sound speed being less than the speed of light. The exact numbers, however,
differ based on how the high core density is matched to the low crustal density,
which is of the order of the nuclear density. If the low density is known to about
twice the nuclear density, one obtains the $\approx 3 M_{\odot}$ upper limit
\citep[see, for example,][]{rhoades_ruffini,Kalogera_1996,lattimer_2012}.
Observational evidences of pulsars obey this limit \citep[see Table 1 of ][]{lattimer_2012}.
The total mass of the GW170817 system, $\approx 2.74 M_{\odot}$, also provides an
observational upper limit. Although the system could have undergone prompt collapse
to form a BH, ejecting some mass prior to it (see Sec.~2.2 of \cite{friedman_2018},
and references therein for a discussion). Some GW template based searches, also,
regard the $3 M_{\odot}$ to be the upper boundary for placing BNS templates
\citep{pycbc_2018}.
Thus, Eq.~(\ref{eq:ml_hasns_label}) is conservative and fundamental
inference about the presence of a NS. However,
we should mention that the presence of compact objects apart from BNS, NSBH, and
BBH which satisfy Eq.~(\ref{eq:ml_hasns_label}) would be included in this inference.
Our inference is only based on the secondary mass, and we do not prejudge the
nature of the object.

For the $\texttt{HasRemnant}$ quantity, to label an injection
as having remnant matter, we use the F18 empirical fit to check for non-vanishing
remnant matter (see Eq.~(4) therein for expression),
\begin{equation}
    M_{\text{rem}}(m_1^{\text{inj}}, m_2^{\text{inj}}, \chi_1^{z {\text{ inj}}}) > 0.
    \label{eq:ml_hasremnant_label}
\end{equation}
The F18 fit requires the compactness of the NS, and hence
an EoS model. For this work, we use the 2H EoS \citep{kyutoku_2010},
which has a maximum NS mass of $2.83 M_{\odot}$. Note that this value is not
to be confused with the value mentioned in Eq.(\ref{eq:ml_hasns_label}), which
is the value considered for the \texttt{HasNS} categorization. The value
$2.83 M_{\odot}$ for $\texttt{HasRemnant}$ comes from the usage of a particular
model EoS. We use the condition in Eq.~(\ref{eq:ml_hasremnant_label}) only for the
injections which have primary mass above the
$2.83 M_{\odot}$ and secondary mass below this value i.e., NSBH systems based on
this EoS. The injections
having both masses less than $2.83 M_{\odot}$ are labeled as having remnant,
while those with both masses above this value are
labeled as not having remnant, based on the assumption that BNS mergers
will always produce some remnant matter, while BBH mergers will will never do so.
The 2H is an unusually stiff EoS resulting in NS radii $\sim 15-16$ km,
but it errs towards larger values of the remnant matter, and therefore
is a conservative choice in the sense of not misclassifying a CBC having remnant
matter as otherwise, due to uncertainty in the EoS. This could be extended to compute
disk masses based on different EoS models reported in the literature, giving each of
them individual astrophysical weight and obtaining an \emph{EoS averaged} disk mass,
and thereby, a $\hasremnant$ after marginalizing over EoS.

We can restrict to the part of the parameter space on
which the classification strongly depends on. We choose the
following set as training features:
\begin{equation}
\label{eq:ml_features}
    \bbeta = \{m_1, m_2, \chi_1^z, \chi_2^z, {\rm SNR}\}\,.
\end{equation}
The reason for using more parameters than those which are used to label
the injections is because the recovered parameters have correlations (see 
Fig.~\ref{fig:ml_parameter_correlation}). For example, the masses are expected
to be positively correlated since the chirp mass is recovered fairly accurately
and is an increasing function of the individual masses. There can also
exist biases in the recovery due to degeneracies in the space of CBC GW signals.
For example, high spin recovery is associated with high mass ratio. Regarding
the choice of the feature set to be used, the masses and primary spins are
natural since they are the intrinsic properties of the binary on which the
source properties depend. As for a detection specific property, we use
the signal to noise ratio, SNR, since it captures the general statistical
uncertainty in the recovered parameters.

With this set, we use the machinery of supervised learning provided by
the \texttt{scikit-learn} library \citep{scikit-learn} to train a binary classifier
based on the search results. Once trained, the classifier outputs a probability
$\hasns$ or $\hasremnant$ given arbitrary but physical values of
$\bbeta$. We tested the performance using two non-parametric algorithms: 
\texttt{KNeighborsClassifier} and \texttt{RandomForestClassifier}, both provided
in the \texttt{scikit-learn} library. We found that the former outperforms the latter
in our case and is used for this study.~\footnote{The nearest-neighbor algorithm
also fits best with the intuition of a map by which the injected parameters, with
the right labels, are \emph{carried} over to the recovered set rather than a
decision tree made by relational operations (which look like ``linear cuts'')
in the parameter space at every \emph{branch} of a decision tree.}
We train it using 11 neighbors -- twice the number of dimensions plus one to break
ties. The collection of parameters of a point-estimate is a point in this parameter
space. To obtain the probability of this point having a secondary
mass $\leq 3 M_{\odot}$ or having some remnant matter based on F18 expression, we use the
nearest neighbors from the training set, weighting them
by the inverse of their distance from the fiducial point,
\begin{eqnarray}
    p(\texttt{HasNS/HasRemnant}) =  
    \frac{\sum_{\tiny{\texttt{HasNS/HasRemnant}}}w_K}{\sum w_K},
\end{eqnarray}
where the numerator (denominator) goes over neighbors that satisfy
Eq.(\ref{eq:ml_hasns_label}, \ref{eq:ml_hasremnant_label}) (all neighbors) of the fiducial point,
and $w_K = 1/d_K$ ($w_K = 1$) for the inverse distance (uniform) weighting. We also
used the \emph{Mahalanobis} metric \citep{mahalanobis} in the space of $\bbeta$
where distance, and therefore, nearest neighbors are determined via,
\begin{equation}
    d_K = (\textbf{x} - \tilde{\textbf{x}})^T \Sigma^{-1}(\textbf{x} - \tilde{\textbf{x}}),
\end{equation}
where $ \tilde{\textbf{x}}$ is the mean and $\Sigma$ is the covariance matrix of the
training set. This is done in the light of handling correlations. We, however, find that the
metric or weighting scheme used does not affect the result
significantly (see Table~\ref{tab:appendix_distance_weighting}). 

\subsection{ROC Curve}
\begin{figure*}[htp]
    \centering
    \includegraphics[width=0.44\textwidth]
	    {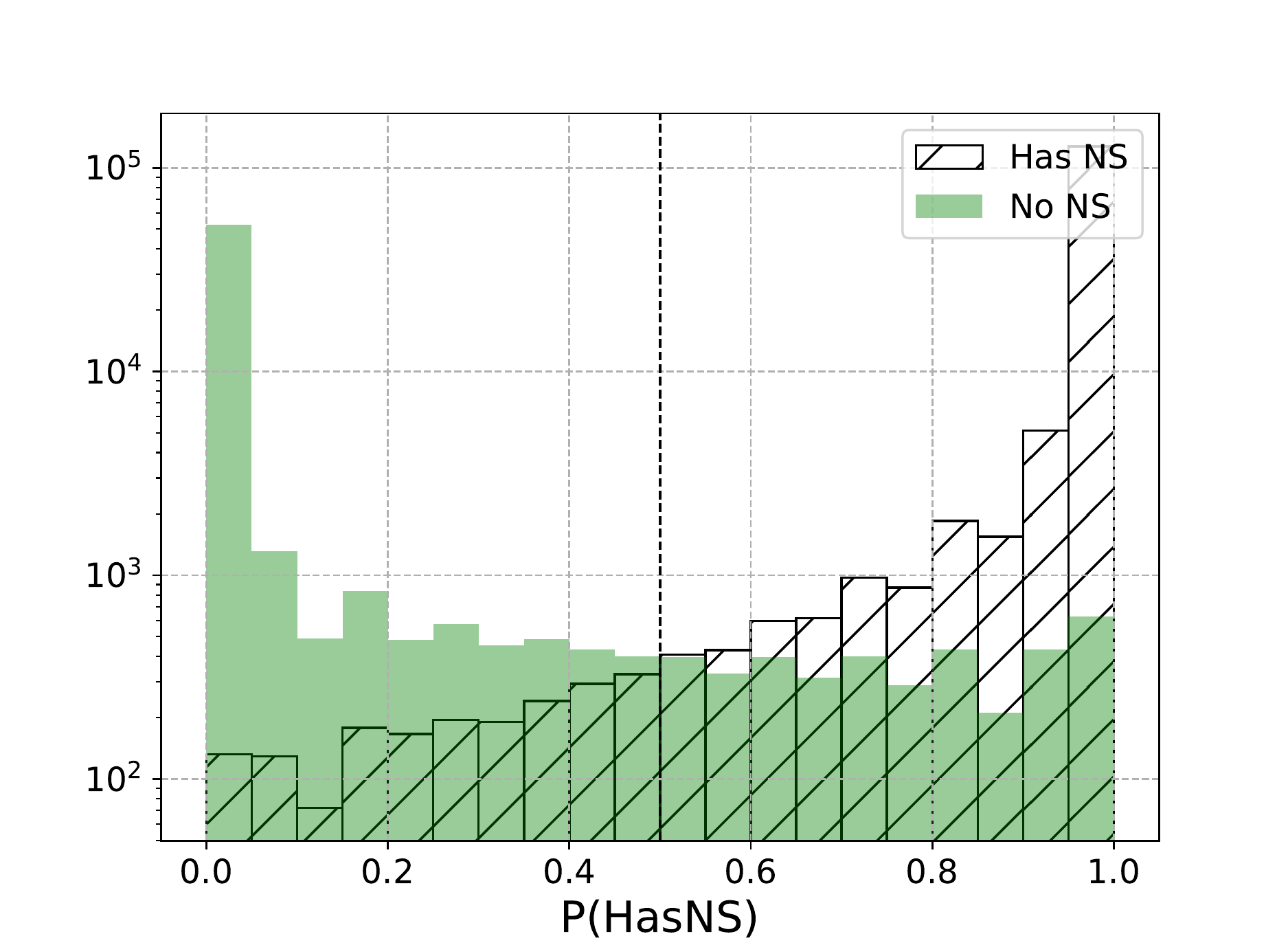}
    \includegraphics[width=0.44\textwidth]
        {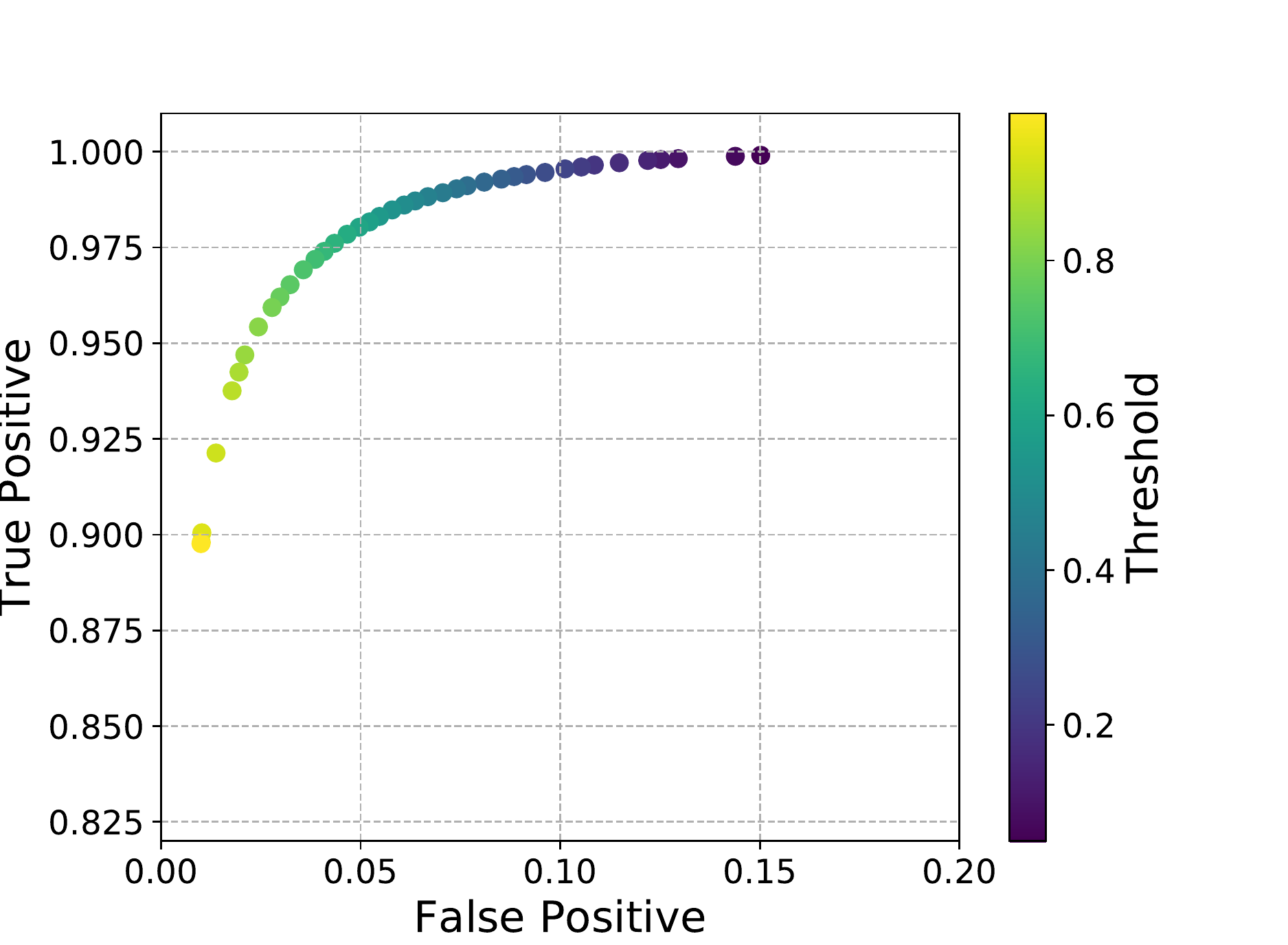}
	\includegraphics[width=0.44\textwidth]
	    {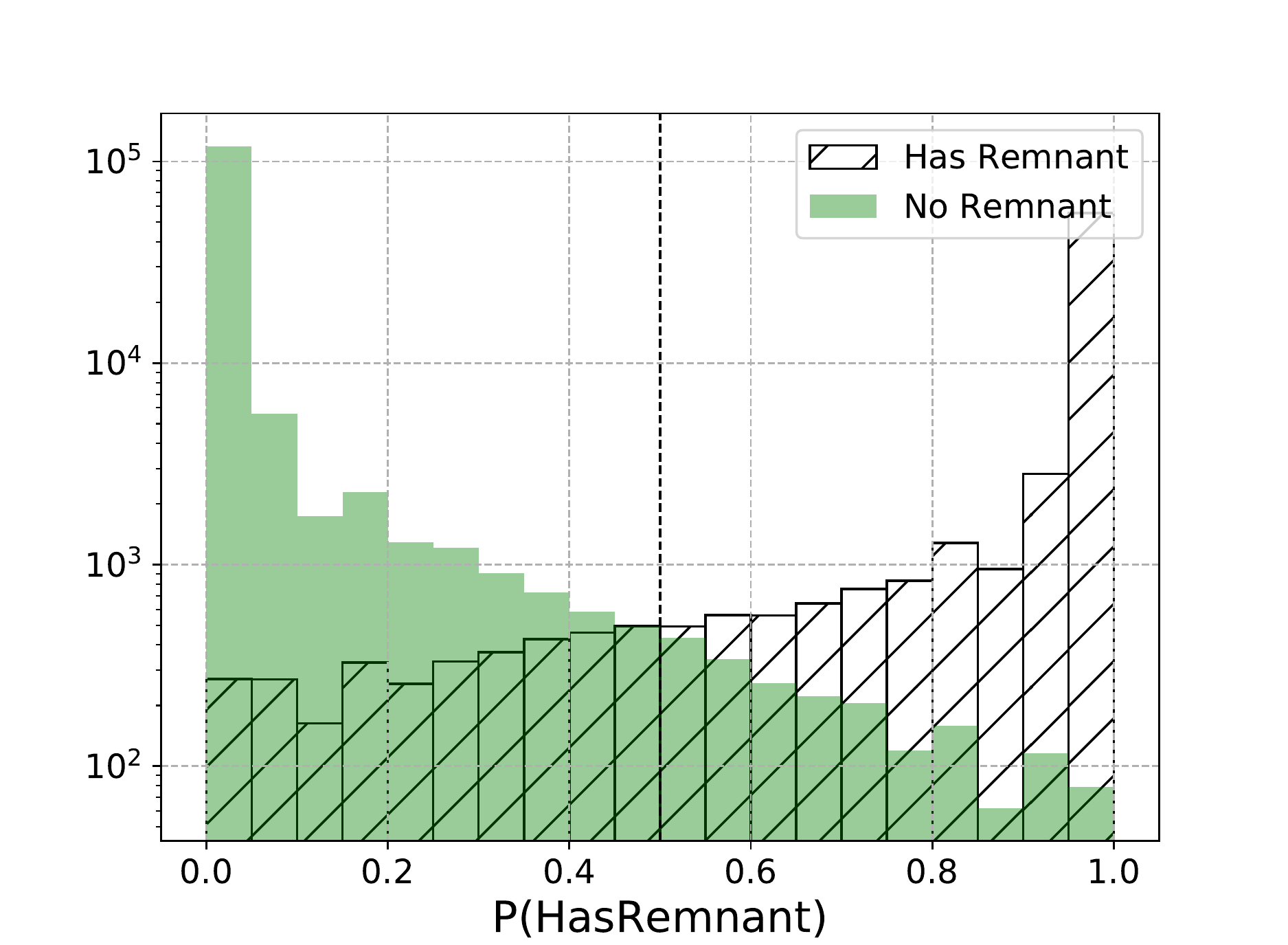}
    \includegraphics[width=0.44\textwidth]
        {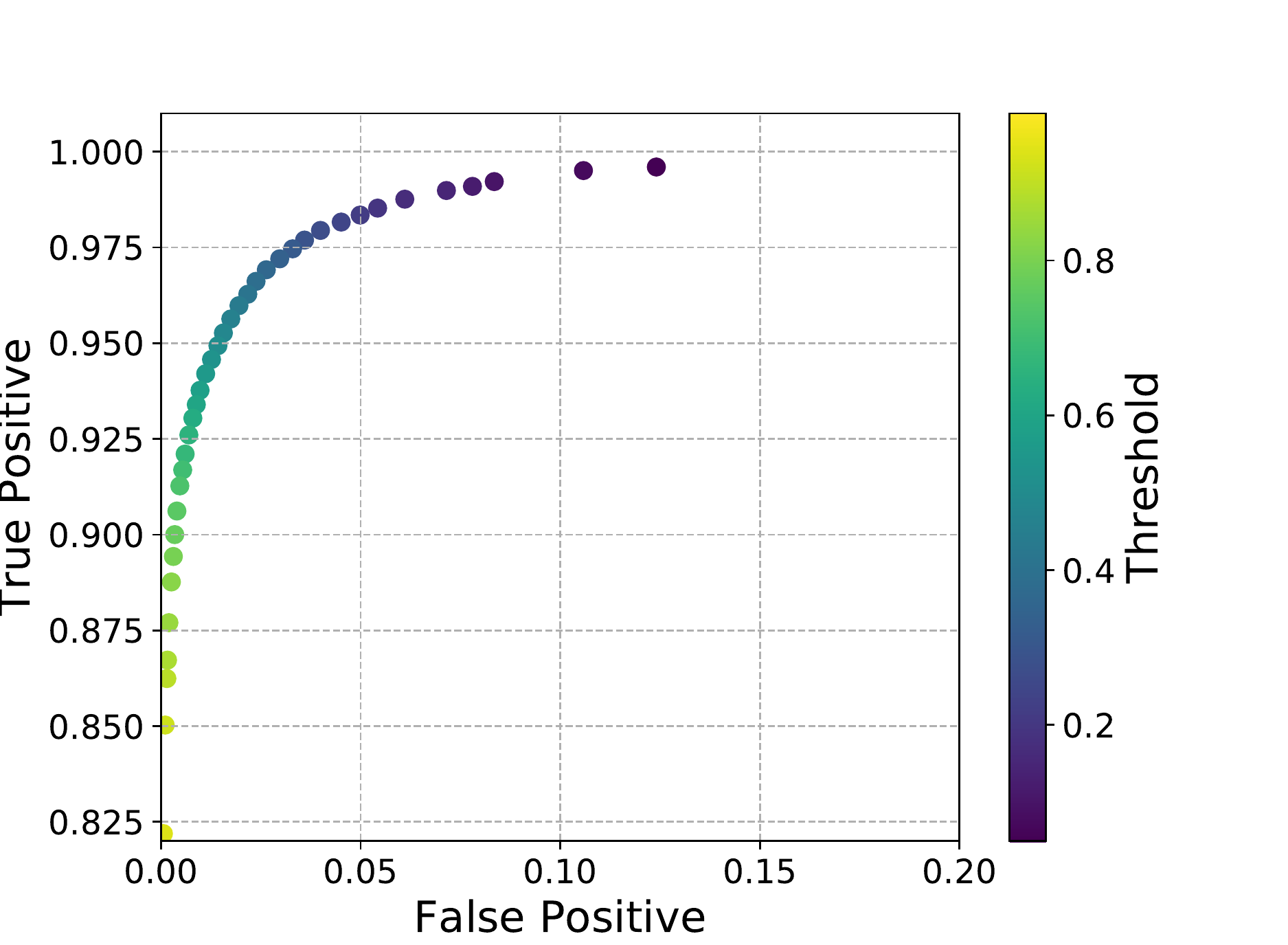}
    \caption{
    This figure shows the 
    receiver operating characteristic curve for the classifier.
    It shows the true-positive against the false positive
    as a function of the threshold to classify binaries as
    having an NS or having remnant matter. \textbf{Top panel}:
    The left figure is a histogram of the $p(\text{HasNS})$
    values for the injections which represented a binary that had an
    NS and for those does that did not. In the limit of perfect performance,
    the values for the former (latter) should be at $p(\text{HasNS}) = 1$
    ($p(\text{HasNS}) = 0$). The true positive and false negative
    performance is decided based on the threshold that is applied
    to make the decision. For example, using the value of 
    $p(\text{HasNS}) = 0.5$ (dot-dashed vertical line) would imply
    that all the values to the right of the line are decided as having
    a NS. While such a decision captures most of the true NS bearing
    binaries, one can notice a small misclassification fraction.
    The right figure shows the fractions as a function of this threshold.
    \textbf{Bottom panel}: Similar plots as the top panel except that the
    values correspond to the binary having remnant matter after merger.
    }
    \label{fig:ml_efficiency_false_alarm}
\end{figure*}
\begin{deluxetable*}{ccccc}
    \tablecaption{The table lists some example values of true positive
    and false positive numbers for changing values of the threshold used
    in Fig.~\ref{fig:ml_efficiency_false_alarm}. The column containing
    threshold values correspond to the colobar in both panels. The true
    positive and false positive values are to be read off based on
    \texttt{HasNS}/\texttt{HasRemnant} case.
    }
    \tablehead{
    \colhead{Threshold} & \colhead{TP(\texttt{HasNS})} &
    \colhead{FP(\texttt{HasNS})} & \colhead{TP(\texttt{HasRemnant})} &
    \colhead{FP(\texttt{HasRemnant})}
    }
    \startdata
	\input{efficiency_false_alarm_values.txt}
    \enddata
\end{deluxetable*}\label{tab:efficiency_false_alarm}
In the case of perfect performance, one expects the trained algorithm to predict
$\hasns = 1$ ($\hasremnant = 1$) from the recovered parameters of the fake
injections which originally had a NS (had remnant matter). On the other hand, in
absence a NS component we also do not expect any remnant matter and hence expect
$p(\texttt{HasNS}/\texttt{HasRemnant}) = 0$.
In order to test the accuracy of the classifier we trained the algorithm on $90\%$
of the dataset and tested it on the remaining $10\%$, cycling the training/testing
combination on the full dataset. The results are shown in
Fig.~\ref{fig:ml_efficiency_false_alarm}. While most of the binaries are correctly
classified as shown in the histogram plot (left panel) for the two quantities,
there is a small fraction which does not end up getting perfect score ($\hasns = 1$).
The choice of threshold value to 
consider a binary suitable for follow-up operations would result in an impurity
fraction. For example, if we use $\hasns \geq 0.5$, shown as a dashed
vertical line in the upper left panel of Fig.~\ref{fig:ml_efficiency_false_alarm},
the contribution of the ``No NS'' histogram to the right of that line constitutes the
false-positive. The variation of the efficiency with the false-positive as a function of
the threshold applied is shown in right panels of Fig.~\ref{fig:ml_efficiency_false_alarm}.
Some example values are listed in Table~\ref{tab:efficiency_false_alarm}.
The threshold could be set depending on the desired efficiency or, alternatively, the
false positive to tolerate. We would like to highlight that the ROC curve
depends on the relative rates of the different astrophysical sources. In this
injection campaign each population has been densely sampled, without considering
the relative rates. However, the current methodology works given an injection
campaign curated based on astrophysical rate estimates of mergers as more
observations are made.

The predictions of a parameter sweep on the $(m_1, m_2)$
values is shown in Fig.~\ref{fig:ml_efficiency_false_alarm}. Considering, the
$\hasns$ plot, a perfect performance of the search would have rendered the
region under the vertical line of $m_2 = 3 M_{\odot}$ as
$\hasns = 1$. In reality, we expect a fuzz around the $m_2 = 3 M_{\odot}$ line,
as shown in the figure. The $\hasremnant$ is behaving as expected with respect to
the increasing spin values, increasing the region having non-vanishing remnant
mass boundary.

%% file: efficiency_false_alarm_values.txt
0.07 & 0.999 & 0.144 & 0.995 & 0.106 \\
0.27 & 0.995 & 0.096 & 0.979 & 0.040 \\
0.51 & 0.986 & 0.061 & 0.949 & 0.014 \\
0.80 & 0.959 & 0.028 & 0.894 & 0.003 \\
0.94 & 0.900 & 0.010 & 0.822 & 0.001 \\

%% file: conclusion.tex
The low-latency inference about the presence of a neutron-star
or post merger remnant matter in a compact binary merger
provides crucial information about whether the binary
will have an EM counterpart, and be
worth following up for the observing partners.
Such time sensitive inference has to be carried out from the
low-latency point-estimate parameters provided by the gravitational wave
realtime search pipelines. However, the point-estimate
masses and spins could be off from the true estimate.
Bayesian parameter estimation provides the best answer to such inference
but it takes $\sim$ hours to $\sim$ days to complete. In order to correct
for such systematics in low-latency, we show the use of supervised
machine learning on the parameter recovery of the GstLAL online search
pipeline from LIGO/Virgo operations. The result is a
binary classifier that is trained based on an injection campaign to
learn such systematics. Once trained, the real time computation on
arbitrary binaries is sub-second. This method is adaptive to the change
of template banks in the low-latency search algorithms provided the
injection campaigns are conducted. Also, it is adaptive to the change
in the noise power spectral density of the interferometer which naturally
manifests in the performance of the search. While we have used a broad training
set for the purposes of this paper, the methodology could be extended
to incorporate astrophysical rates by curating injection campaigns
based on our knowledge of the rates of binary mergers.

%% file: acknowledgements.tex
This work was supported by NSF grant no. PHY-1700765 and PHY-1912649.
D.C. acknowledges the use of computing resources of the LIGO Data Grid
and facilities provided by Leonard E. Parker Center for Gravitation,
Cosmology and Astrophysics at University of Wisconsin-Milwaukee.
D.C. would like to thank Jolien Creighton, Siddharth Mohite, and
Duncan Meacher for helpful discussions. The authors would like to thank
the anonymous referee for helpful comments.

\software{
scikit-learn \citep{scikit-learn}, 
Matplotlib \citep{Hunter:2007}, scipy \citep{scipy}, numpy \citep{numpy},
pandas \citep{mckinney-proc-scipy-2010}, jupyter (\url{https://jupyter.org/}),
SQLAlchemy (\url{https://www.sqlalchemy.org/}).
}

%% file: appendix.tex
\section{Parameter sweep showing variation with SNR}
\begin{figure*}[htp]
    \begin{minipage}{0.49\textwidth}
    \begin{center}
    \includegraphics[width=1.0\textwidth]
        {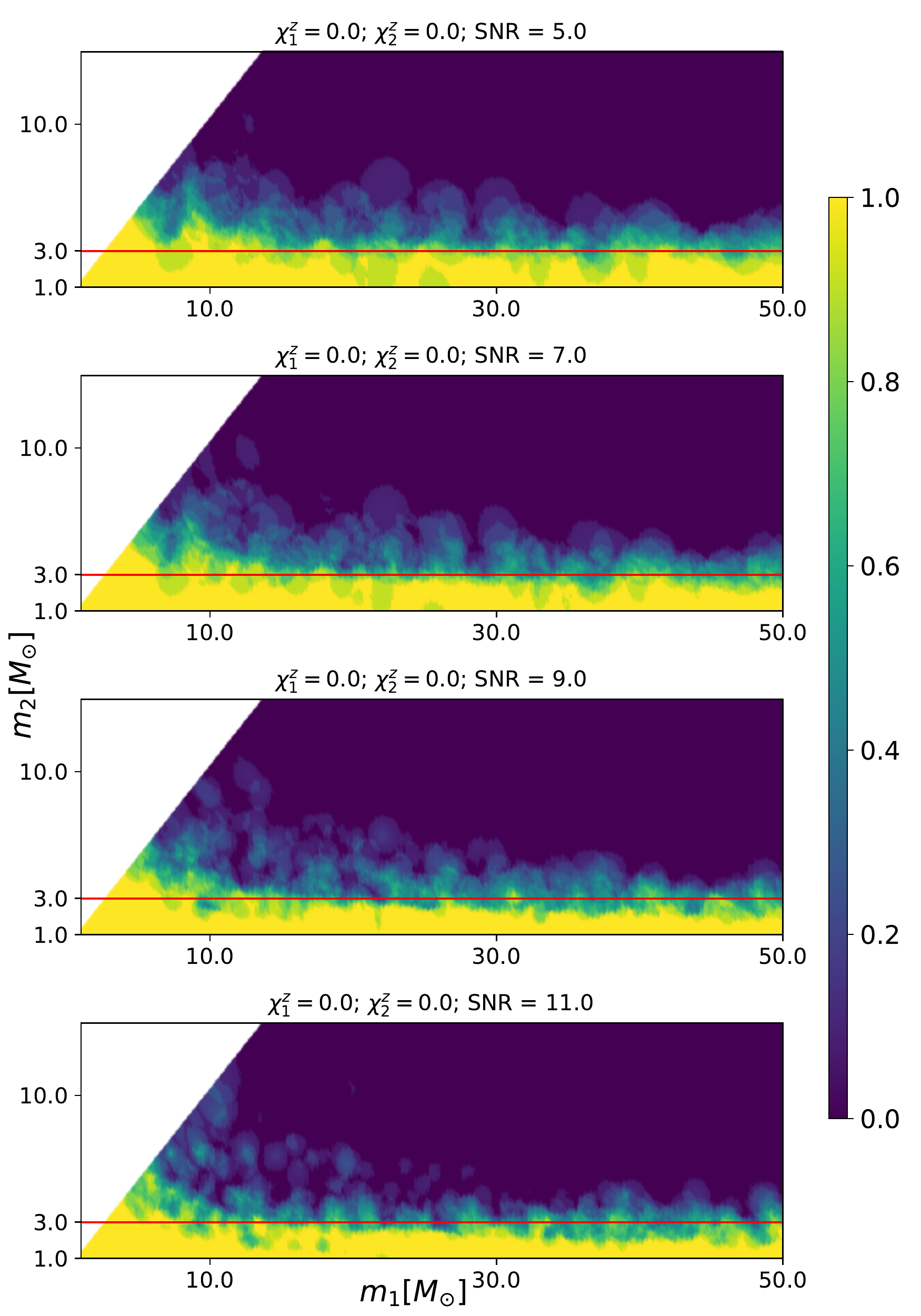} \\
    $\hasns$
    \end{center}
    \end{minipage}
    \begin{minipage}{0.49\textwidth}
    \begin{center}
        \includegraphics[width=1.0\textwidth]
        {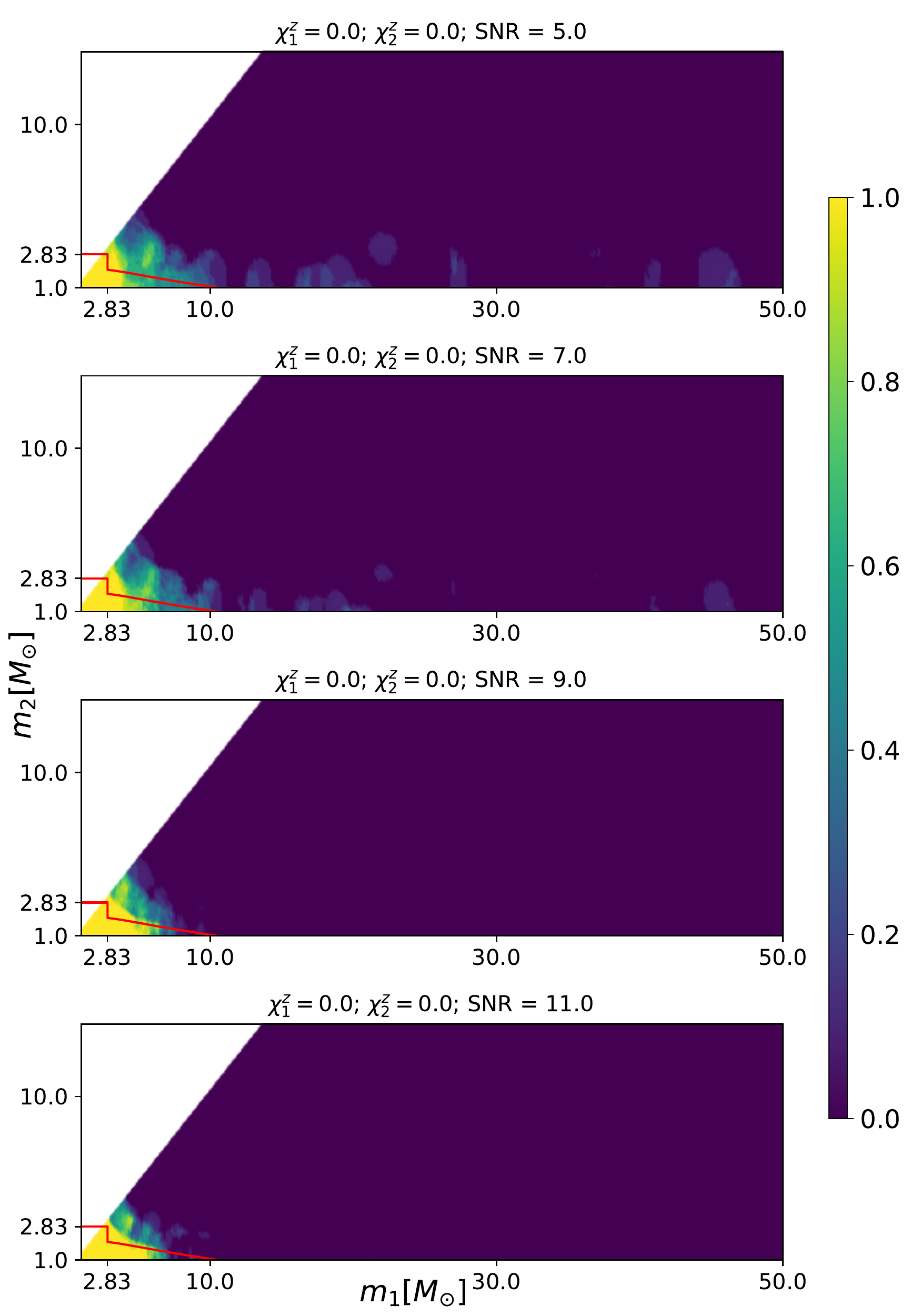} \\
    $\hasremnant$
    \end{center}
    \end{minipage}
    \caption{
    This figure is an extension of Fig.~\ref{fig:ml_parameter_sweep}.
    Here we see the behavior of the predictions from the binary
    classifiers as the signal to noise (SNR) of recovery increases.
    \textbf{Left panel:} Variation in $\hasns$ with SNR. \textbf{Right panel:}
    Variation in $\hasremnant$ with SNR.
    }
    \label{fig:ml_parameter_sweep_snr_variation}
\end{figure*}

In this section, we make in extension of the parameter sweep results
shown in Fig.~\ref{fig:ml_parameter_sweep}. Here we sweep over the
$(m_1, m_2)$ values but keep the values of the spins fixed, only varying
the signal-to-noise (SNR). The result is shown in
Fig.~\ref{fig:ml_parameter_sweep_snr_variation}. It is expected that the
uncertainty in the recovered parameter should decrease with the increase
in SNR which manifests as a decrease in the \emph{fuzzy} region separating
the \emph{bright} ($\hasns = 1$/$\hasremnant = 1$) and \emph{dark}
($\hasns = 0$/$\hasremnant = 0$) regions.

\section{GstLAL Injection sets}\label{appendix:injection_sets}
\begin{deluxetable*}{lll}
    \label{tab:appendix_injection_sets}
    \tablecaption{The table contains the calender times for two detector (H1L1)
    chunks of LIGO O2 data. \replaced{The GstLAL search performed several injection campaigns
    in this data. We consider the broad injection campaign used in spacetime
    volume sensitivity analysis in \cite{ligo_catalog_paper}.}{We consider the
    injections performed by the GstLAL search in these durations for this
    study. The timeseries is available in \url{https://www.gw-openscience.org/data/}}}
    \tablehead{
    \colhead{GstLAL chunk} & \colhead{Start date} &
    \colhead{End date}
    }
    \startdata
    Chunk 02 & \texttt{Wed Nov 30 16:00:00 GMT 2016} & \texttt{Fri Dec 23 00:00:00 GMT 2016} \\
    Chunk 03 & \texttt{Wed Jan 04 00:00:00 GMT 2017} & \texttt{Sun Jan 22 08:00:00 GMT 2017} \\
    Chunk 04 & \texttt{Sun Jan 22 08:00:00 GMT 2017} & \texttt{Fri Feb 03 16:20:00 GMT 2017} \\
    Chunk 05 & \texttt{Fri Feb 03 16:20:00 GMT 2017} & \texttt{Sun Feb 12 15:30:00 GMT 2017} \\
    Chunk 06 & \texttt{Sun Feb 12 15:30:00 GMT 2017} & \texttt{Mon Feb 20 13:30:00 GMT 2017} \\
    Chunk 07 & \texttt{Mon Feb 20 13:30:00 GMT 2017} & \texttt{Tue Feb 28 16:30:00 GMT 2017} \\
	Chunk 08 & \texttt{Tue Feb 28 16:30:00 GMT 2017} & \texttt{Fri Mar 10 13:35:00 GMT 2017} \\
    Chunk 09 & \texttt{Fri Mar 10 13:35:00 GMT 2017} & \texttt{Sat Mar 18 20:00:00 GMT 2017} \\
	Chunk 10 & \texttt{Sat Mar 18 20:00:00 GMT 2017} & \texttt{Mon Mar 27 12:00:00 GMT 2017} \\
    Chunk 11 & \texttt{Mon Mar 27 12:00:00 GMT 2017} & \texttt{Tue Apr 04 16:00:00 GMT 2017} \\
    Chunk 12 & \texttt{Tue Apr 04 16:00:00 GMT 2017} & \texttt{Fri Apr 14 21:25:00 GMT 2017} \\
    Chunk 13 & \texttt{Fri Apr 14 21:25:00 GMT 2017} & \texttt{Sun Apr 23 04:00:00 GMT 2017} \\
    Chunk 14 & \texttt{Sun Apr 23 04:00:00 GMT 2017} & \texttt{Mon May 08 16:00:00 GMT 2017} \\
    Chunk 15 & \texttt{Fri May 26 06:00:00 GMT 2017} & \texttt{Sun Jun 18 18:30:00 GMT 2017} \\
    Chunk 16 & \texttt{Sun Jun 18 18:30:00 GMT 2017} & \texttt{Fri Jun 30 02:30:00 GMT 2017} \\
    Chunk 17 & \texttt{Fri Jun 30 02:30:00 GMT 2017} & \texttt{Sat Jul 15 00:00:00 GMT 2017} \\
    Chunk 18 & \texttt{Sat Jul 15 00:00:00 GMT 2017} & \texttt{Thu Jul 27 19:00:00 GMT 2017} \\
    \enddata
\end{deluxetable*}
In this section, we report the calender dates for the \replaced{injection sets}{data
chunks} used in this study. These are tabulated in Table~\ref{tab:appendix_injection_sets}.
The chunks cover most of the duration of the observing run, although they may not be contiguous
corresponding to break in the observing run. Three detector injections were performed
about the last $\sim 1$ month of the second observing run. Thus, their length and hence
the missed found injections are smaller in number. As a future work, we plan to re-analyze
the performance of the classifier based on injection campaigns in the third observing
run as they are performed.
\FloatBarrier